\documentclass[preprint,authoryear,12pt]{elsarticle}

\usepackage{amssymb}
\usepackage{xcolor}
\usepackage{enumitem}
\usepackage{apalike}
\usepackage[round]{natbib}
\usepackage{url}

\journal{Computers and Education}

\begin{document}

\begin{frontmatter}

\title{AI Hallucination from Students' Perspective: A Thematic Analysis}

\author[inst1]{Abdulhadi Shoufan\corref{cor1}}

\affiliation[inst1]{
  organization={Department of Computer and Information Engineering, Center for Cyber Physical Systems, Khalifa University},
  city={Abu Dhabi},
  country={United Arab Emirates}
}
\ead{abdulhadi.shoufan@ku.ac.ae}

\author[inst2]{Ahmad-Azmi-Abdelhamid Esmaeil}

\affiliation[inst2]{
  organization={Department of Psychology, University Malaysia Sabah},
  city={Kota Kinabalu},
  country={Malaysia}
}

\cortext[cor1]{Please address correspondence to Abdulhadi Shoufan}

% \author[inst1]{Abdulhadi Shoufan\corref{cor1}}

% \affiliation[inst1] {organization={Department of Computer and Information Engineering, Center for Cyber Physical Systems, Khalifa University},%Department and Organization
%             city={Abu Dhabi},
%             country={United Arab Emirates}}
% \ead{abdulhadi.shoufan@ku.ac.ae}

% \author[inst2]{Ahmad-Azmi-Abdelhamid Esmaeil}

% \affiliation[inst2]{organization={Department of Psychology, University Malaysia Sabah},%Department and Organization
%             city={Kota Kinabalu},
%             country={Malysia}

% \cortext[cor1]{Please address correspondence to Abdulhadi Shoufan}

\begin{abstract}
% 2025-12-18

As students increasingly rely on large language models, hallucinations pose a growing threat to learning. To mitigate this, AI literacy must expand beyond prompt engineering to address how students should detect and respond to LLM hallucinations. To support this, we need to understand how students experience hallucinations, how they detect them, and why they believe they occur. To investigate these questions, we asked university students three open-ended questions about their experiences with AI hallucinations, their detection strategies, and their mental models of why hallucinations occur. Sixty-three students responded to the survey. Thematic analysis of their responses revealed that reported hallucination issues primarily relate to incorrect or fabricated citations, false information, overconfident but misleading responses, poor adherence to prompts, persistence in incorrect answers, and sycophancy. To detect hallucinations, students rely either on intuitive judgment or on active verification strategies, such as cross-checking with external sources or re-prompting the model. Students’ explanations for why hallucinations occur reflected several mental models, including notable misconceptions. Many described AI as a research engine that fabricates information when it cannot locate an answer in its “database.” Others attributed hallucinations to issues with training data, inadequate prompting, or the model’s inability to understand or verify information. These findings illuminate vulnerabilities in AI-supported learning and highlight the need for explicit instruction in verification protocols, accurate mental models of generative AI, and awareness of behaviors such as sycophancy and confident delivery that obscure inaccuracy. The study contributes empirical evidence for integrating hallucination awareness and mitigation into AI literacy curricula.
\end{abstract}

\begin{keyword}
Large Language Models \sep education \sep hallucination  \sep AI literacy
\end{keyword}

\end{frontmatter}

\newpage
\section{Introduction}
\label{sec:introduction}

Large language models (LLMs) such as ChatGPT have become integral to university students' academic work. Recent studies document that nearly 90\% of students use AI tools for assignments and coursework~\citep{Archana2025,eric2025action,freeman2025studentAI}, with this pattern extending globally. For example, one survey across 16 countries found 86\% student adoption~\citep{digital2024global}. 

However, the integration of LLMs into academic settings introduces significant challenges, particularly concerning content accuracy. One major issue is the phenomenon known as Hallucinations, which is defined as the generation of content that appears fluent and authoritative but is factually incorrect, fabricated, or unsupported by evidence~\citep{Huang2025}. Hallucinations can manifest as fabricated citations with non-existent authors, false historical facts or dates, invented statistics, incorrect explanations of complex concepts, and flawed code presented with apparent confidence~\citep{Sun2024}. These hallucinations arise because LLMs are trained to predict next word sequences rather than retrieve verified facts. They lack mechanisms to distinguish between learned patterns and plausible-sounding fabrications~\citep{Bhattacharya2024}.

Research on LLM hallucinations has increased rapidly, primarily focusing on computational detection and mitigation strategies. Different studies have developed assessment frameworks to measure hallucination rates across specialized domains such as medical diagnosis~\citep{Hazra2025,Ye2025} and legal reasoning~\citep{Dahl2024}, with automated detection systems using semantic entropy and fact-checking pipelines~\citep{Farquhar2024,Heo2025}. Besides that, other studies presented different Technical approaches to reduce hallucinations. These include retrieval-augmented generation (RAG) systems that ground model outputs in verified external sources~\citep{Mohammed2025,Qu2025}, contrastive decoding algorithms that adjust probability distributions to suppress fabricated content~\citep{Hongying2025,Yu2025}, and knowledge graph integration to provide structured factual constraints~\citep{Lavrinovics2025}.  

However, these technical approaches focus on system-level detection and mitigation rather than understanding how users, such as students, encounter and respond to hallucinations in practice. Students' perceptions and experiences are highly relevant for education, as research demonstrates that perceptions significantly influence engagement and learning outcomes. Students' course perceptions predict their engagement, which in turn predicts their learning~\citep{Jones2019}. Also, perceived limitations or capabilities of learning tools affect psychological vulnerability, class engagement, and academic performance~\citep{Muenks2020}. 

Research on student perceptions and experiences with AI in education has grown substantially. This research documents diverse aspects of student engagement with generative AI: ethical implications and gaps in awareness of institutional policies regarding GenAI use~\citep{Morari2025}, adoption facilitators and barriers across professional programs including nursing, pharmacy, radiology, and dentistry~\citep{Alexander2026,Ma2025,Sarangi2024,Sarhan2025}, integration strategies for writing and research tasks in general education~\citep{Black2025}, lived experiences navigating immediacy, equity, and integrity concerns~\citep{Holland2024}, and privacy implications of AI systems~\citep{Wang2025}. Additional studies examine AI literacy levels and attitudes toward AI in clinical contexts~\citep{Si2025}, perceptions of AI as a collaborative teammate~\citep{Marrone2024}, and domain-specific applications such as clinical case companions~\citep{Chastain2025} and public speaking training~\citep{Chen2025}.

However, there is limited research that addresses user or student experiences with hallucinations specifically. ~\citet{Massenon2025}  analyzed three million mobile app reviews and found that 1.75\% of AI error reports concerned hallucinations, with factual incorrectness (38\%), nonsensical outputs (25\%), and fabricated information (15\%) as the most common types. ~\citet{Rapp2025}  examined how people react to nonsensical responses from LLMs and found that novice users often see these unpredictable outputs as signs of independent behavior rather than simple mistakes. As a result, they tend to anthropomorphize the system and may feel unsettled by its responses. However, neither study examines students in educational contexts. Given the increasing usage of LLMs by students, it is important to understand their experiences with and mental models of hallucinations. This study investigates this aspect by answering the following three research questions: 

RQ1: What types of hallucinations do students report encountering?

RQ2: What strategies do students use to identify hallucinations?

RQ3: How do students explain the phenomenon of LLM hallucination?

\newpage
\section{Methodology}
\label{sec:methodology}

\paragraph{Study design}

This study employed a qualitative design to explore university students' encounters with and conceptualizations of LLM hallucinations in educational contexts. Rather than relying on predetermined categories or researcher assumptions about hallucination experiences, we used open-ended questions to allow students to describe their authentic experiences in their own words. This approach enabled the discovery of the types of hallucinations students actually encounter, the strategies they spontaneously employ to detect fabricated information, and their intuitive theories about why hallucinations occur. 
The current study study was approved by the Research Ethics Committee at Khalifa University.

\paragraph{Participants and context}

The participants in this study were 63 senior students in a computer engineering program in the Fall 2025 semester. This program includes courses where students use LLMs for various academic tasks, including code generation, debugging, algorithm explanation, concept clarification, research assistance, and technical writing. 

\paragraph{Open-ended questions}

Students were asked to respond to three open-ended questions through a Moodle questionnaire:

\begin{enumerate}
    \item Describe one or more relevant examples where you observed LLM hallucination!
    \item Describe how you identify LLM hallucinations?
    \item Do you know why LLMs hallucinate? Discuss!
\end{enumerate}

\paragraph{Thematic analysis of students' responses}

Students' responses to the open-ended questions were analyzed using thematic analysis, a method for identifying, analyzing, and reporting patterns within qualitative data~\citep{Braun2006}. We used Taguette, a free and open-source qualitative data analysis tool, to support the coding process. This application enables systematic encoding of text segments, making it easier to identify patterns and themes across responses. The analysis proceeded through multiple stages:

\textbf{Initial coding:} Responses were read thoroughly, and meaningful text segments were assigned descriptive codes using Taguette. A text segment could be a sentence, part of a sentence, or multiple sentences that expressed a coherent idea. For example, when a student mentioned "ChatGPT cited papers that don't exist," this was coded as "fabricated citations." When segments contained multiple distinct concepts, multiple codes were applied. Taguette's tagging functionality allowed us to highlight text segments and assign one or multiple codes to each segment.

\textbf{Code development:} Codes emerged iteratively throughout the analysis. When a new concept appeared in the data, a new code was created. Similar concepts were assigned the same code to enable pattern identification. This iterative process continued across all responses to all three research questions. 

\textbf{Coding reliability:} The first author initially coded all responses according to the emergent coding scheme. The coded data were then exported to Excel for verification. The second researcher reviewed each coded segment and indicated agreement (1) or disagreement (0) with the assigned codes. The coding agreement was 86.93\% for RQ1 (types of hallucinations encountered), 93.16\% for RQ2 (detection strategies used), and 92.72\% for RQ3 (explanations of hallucination causes). Disagreements were resolved through discussion until consensus was reached. Through this consensus process, the final coding scheme consisted of 42 codes for RQ1, 42 codes for RQ2, and 30 codes for RQ3.

\textbf{Theme building:} After achieving consensus on codes, these were reviewed systematically to identify higher-order patterns. Similar codes were grouped into themes representing broader categories. For example, codes for "fabricated citations," "non-existent sources," and "wrong authors" were grouped into a theme about citation-related hallucinations. This theme-building process involved multiple rounds of comparison and refinement, resulting in 4 themes for RQ1, 2 themes for RQ2, and 5 themes for RQ3. The final coding scheme and theme structure were documented for transparency and replicability.

\newpage

\section{Results}
\label{sec-results}

\subsection{RQ1 (What types of hallucinations do students report encountering?)}
\label{sec-results-RQ1}

\begin{figure}[h]
	\centering
	\includegraphics[width=12 cm]{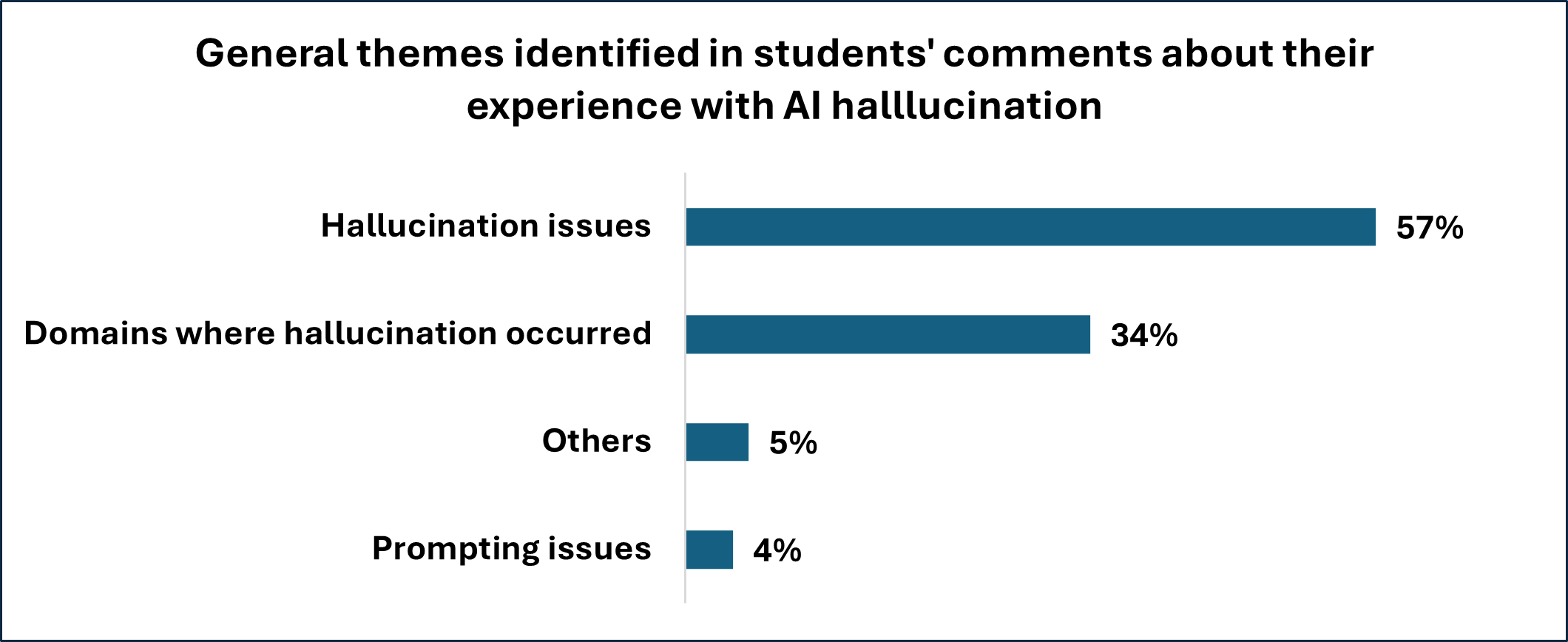}
	\caption{A total of 152 students’ comments about their experiences with AI hallucination were coded into four main themes.}
	\label{fig-RQ1-main-themes}
\end{figure}

Fifty-eight of the sixty-three students responded meaningfully to the first question and commented on their experiences with AI hallucination. We coded a total of 152 comments using 42 emergent tags, which were grouped into four main themes, as shown in Figure~\ref{fig-RQ1-main-themes}. Most comments (91\%) described hallucination issues or referred to the domains in which hallucinations occurred. These will be described in the following two subsections in detail.

A small number of comments (4\%) attributed hallucinations to prompting issues, such as prompts that were too long, incomplete, or insufficiently specific. Half of the comments in the \textit{Others} theme reflected students’ perceptions that newer models hallucinate less, while the other half expressed emotional reactions to hallucination, e.g. e.g., \emph{"it was disappointing and quite creepy"}, \emph{"which is very frustrating"}, \emph{"More pressingly, the LLM would occasionally hallucinate"}, and \emph{"quite astonishing for such a simple mistake to be done by an AI"}.

\subsubsection{Hallucination issues}
\label{sec-results-RQ1-hal-issues}

\begin{figure}[h]
	\centering
	\includegraphics[width=12 cm]{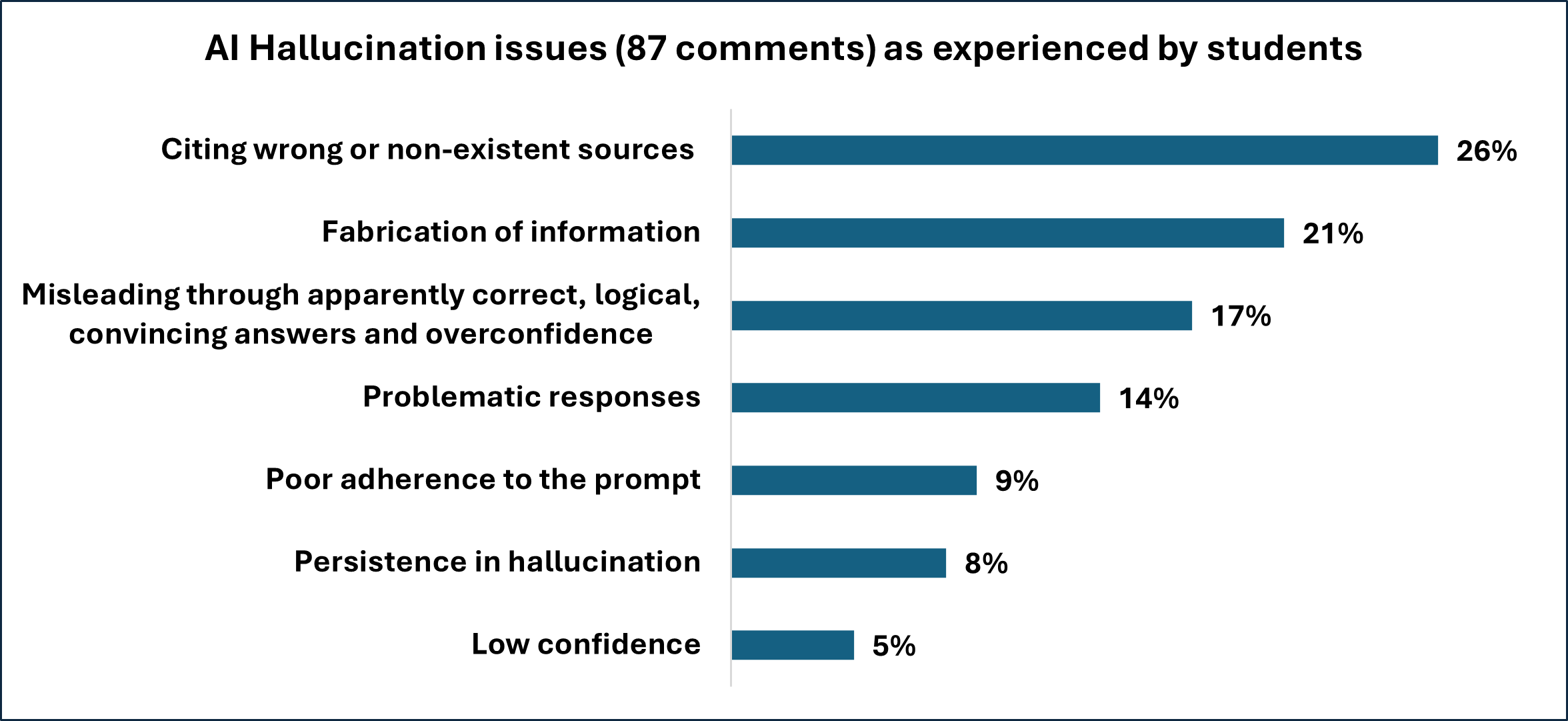}
	\caption{Students' comments about AI hallucination issues were grouped into seven subthemes.}
	\label{fig-RQ1-hallucination-issues}
\end{figure}

The eighty-seven comments related to hallucination issues were organized into seven subthemes, as shown in Figure \ref{fig-RQ1-hallucination-issues}. Nearly one-quarter of these comments referred to instances of citing incorrect or non-existent sources. For example, one student stated, \emph{"One time, I asked it to give me a journal paper about a certain topic, and the journal paper was a pdf of a university shared by students"}. Another student commented, \emph{"It gave me some insights and when I asked about the sources, it sent me fake ones that don’t exist.}

Fabrication of information was the second most frequently mentioned issue. Although citing non-existent sources can be considered a form of fabrication, this aspect was placed in a separate category due to its particular relevance. One student explained, \emph{"when I asked an AI model to give me the biography of a scientist. It gave me a detailed answer with dates, education, and achievements but when I checked online, some of the facts were wrong or completely made up."} Another student reported, \emph{"when asked to summarize a scientific paper, the model included fake statistics and references that did not appear in the original text."}

Several students highlighted that the model’s confidence and apparent clarity and correctness could mislead users into believing that its answers were accurate. One student commented, \emph{"I’ve also seen it give Arduino code that looks correct but doesn’t actually work when tested. So basically, it can sound very convincing even when the information is wrong."}

Fourteen percent of students’ comments pointed to various issues in the generated responses, such as being incomplete, too general, or irrelevant. One student noted \emph{"doesn't provide the full answer, especially when we ask a specific question"}. Another student stated, \emph{"giving very generic responses that did not relate to the specific product that I asked for at all."}

Some students reported that the AI model can fail to understand or strictly adhere to the prompt, which could lead to hallucinations. For instance, the model occasionally ignored files attached to the prompt. One student shared, %\emph{"Another example occurred when I asked the model to summarize a scientific paper I uploaded. Instead of sticking to the actual content, it invented details about the study’s results and conclusions that were not in the text."}. 
\emph{"I ask something and it gives another thing."} 
Another issue is when the model expands the prompt by introducing assumptions that students had not made. One student reported, \emph{"it could provide an answer by filling in assumptions that you as the prompter have not included."}

Another aspect of hallucination reported by students was persistence, meaning that the model sometimes became trapped in a state where repeated prompting did not help redirect it toward correct answers. One student reported, \emph{"sometimes falls into a loop and couldn't find the correct and accurate solution. It tells you something, and you do and then when it doesn't work, it goes back to the beginning and hence looping in the solution."} Another student commented, \emph{"gave me an incorrect answer and kept digging deeper."} 

Ironically, while the model’s overconfidence could mislead students, as described above, some students observed that the model occasionally displayed low confidence by contradicting itself, apologizing, or acting sycophantically. One student remarked, \emph{"In this case, however, the LLM would jump from pin number to pin number contradicting itself constantly."}. Another student stated, \emph{"when you give it the answers with the steps it resolves it again with the way you gave even though sometimes the solution you provided is wrong but it doesnt want to prove you wrong."}

\subsubsection{Hallucinaion domains}
\label{sec-results-RQ1-domains}
\begin{figure}[h]
	\centering
	\includegraphics[width=12 cm]{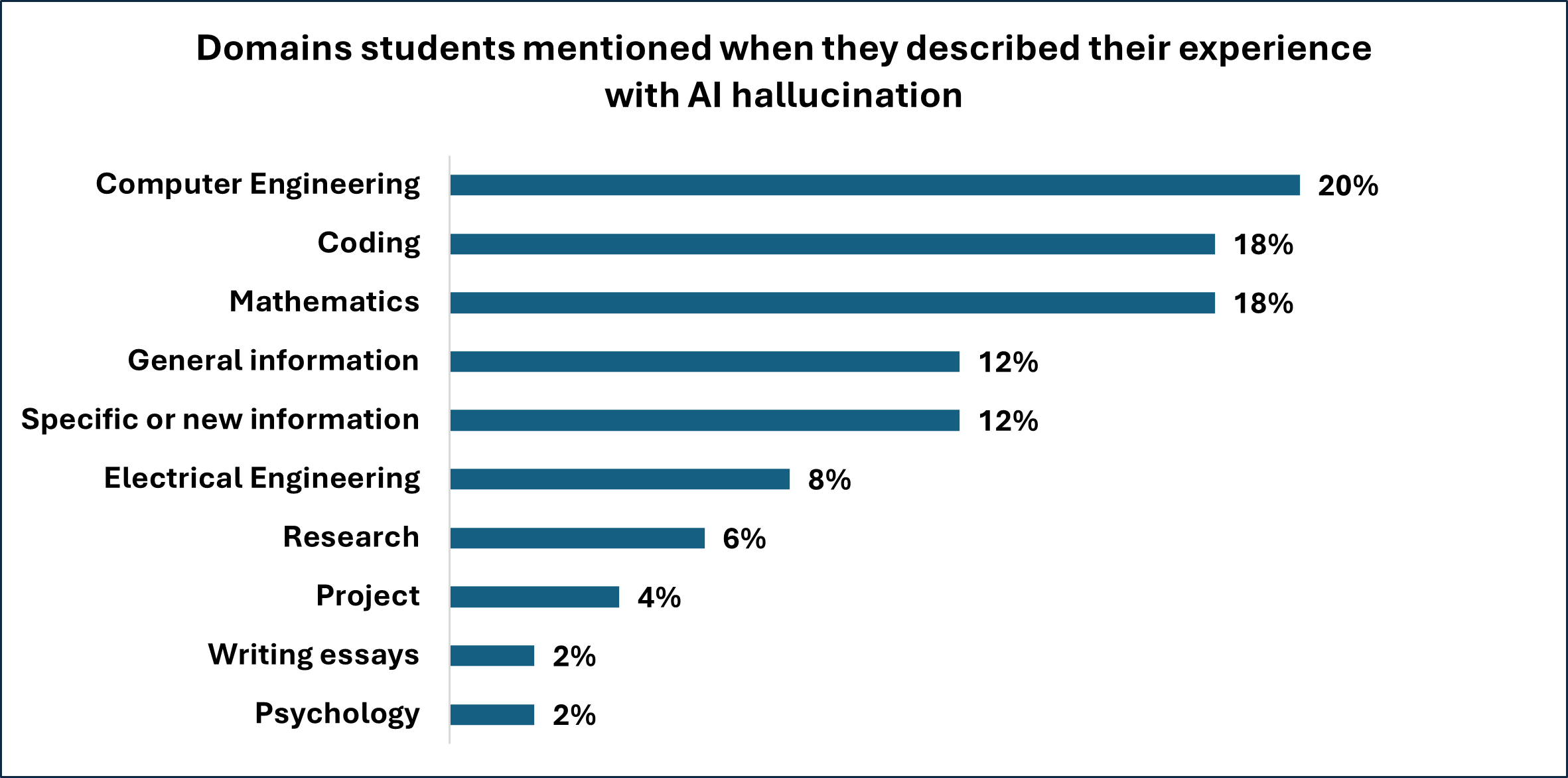}
	\caption{Domains students mentioned when they reported their experience with AI hallucination.}
	\label{fig-RQ1-domains}
\end{figure}

Figure \ref{fig-RQ1-domains} summarizes the domains students mentioned while describing their experiences with AI hallucination. As expected, most of these domains are related to the students’ major, computer engineering, and associated fields such as programming, mathematics, and electrical engineering. About one-quarter of the comments referred to hallucinations that occurred when students prompted the model for general information or for specific, unfamiliar topics. Ten percent of the comments indicated that students experienced AI hallucinations while using it to complete a project or conduct research.

\newpage

\subsection{RQ2 (What strategies do students use to identify hallucinations?)}
\label{sec-results-RQ2}

\begin{figure}[h]
	\centering
	\includegraphics[width=10 cm]{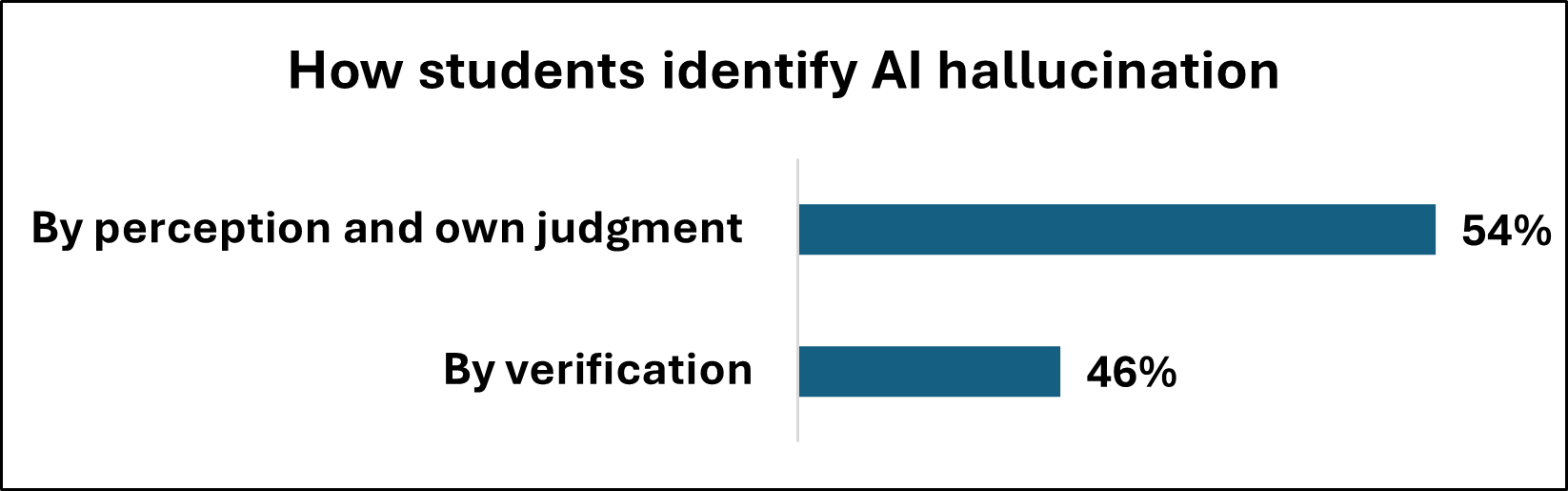}
	\caption{Students identify AI hallucination either by their perception and direct judgment or by using some kind of verification.}
	\label{fig-RQ2-main-categories}
\end{figure}

Fifty-four of the sixty-three students answered the second question, which asked how they identify AI hallucinations. Seven students did not provide any response, and two students explicitly stated that they could not identify hallucinations. Among the collected responses, nine comments were deemed unhelpful because they were unrelated to the question. For instance, one student wrote, \emph{"The hallucination is there and it is unavoidable until now."} Excluding these, 96 comments were considered useful and were grouped into two main themes, as illustrated in Figure \ref{fig-RQ2-main-categories}.

\subsubsection{Students' perceptions of AI hallucination}
\label{sec-results-RQ2-perception}

\begin{figure}[h]
	\centering
	\includegraphics[width=12 cm]{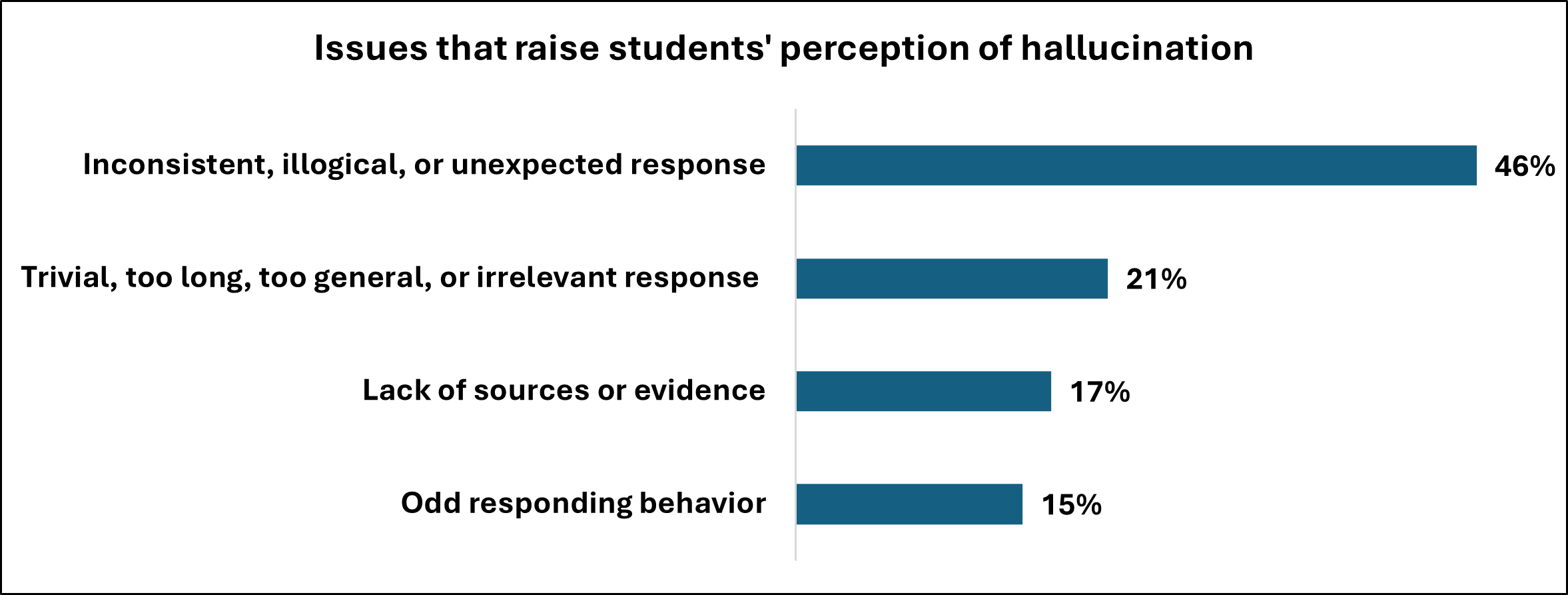}
	\caption{Issues that raise students' perception of hallucination
}
	\label{fig-RQ2-perception}
\end{figure}

Figure \ref{fig-RQ2-perception} summarizes how students perceive an AI response as a hallucination. Slightly less than half of the comments indicate that students identify hallucinations when the AI response appears inconsistent, illogical, or unexpected. Examples of student comments include: \emph{"when the answer is illogical and clearly incorrect,"}  \emph{"when it talks nonsense,"}  \emph{"the answer doesn’t make sense,"} and \emph{"When i see an answer that looks really abroad from expected answer even though you dont know how to solve it."}

Another aspect that helps students recognize hallucinations is the coverage of the AI response. Nearly one-fifth of the comments indicate that students perceive trivial, too detailed, excessively long, too general, irrelevant, or unrelated responses as signs of AI hallucination. Examples of student comments include: \emph{"over explaining something simple, using big words to make it seem intelligent,"} \emph{"When the explanation feels too general,"}  \emph{"emphasizing on trivial matters,"}  \emph{"the answer isn’t related to my question"}, \emph{"Making the user to believe false information by adding extra info trying to show that its spitting facts."}  

Furthermore, some students feel that the model hallucinates when it does not provide sources or evidence to support its response. Examples of student comments include: \emph{"Lack of resources"} and \emph{"A hallucinated answer often sounds confident but doesn’t provide clear evidence, citations, or logic to support it."} 

Finally, some students perceive hallucination when the model behaves oddly, for example, when it produces responses \emph{"based on prior conversations,"} \emph{"tends to repeat itself"}, \emph{"agrees with everything you say"}, or loses context \emph{"when the chat has been too long."}

\subsubsection{Detecting hallucination by verification}
\label{sec-results-RQ2-verification}
\begin{figure}[h]
	\centering
	\includegraphics[width=12 cm]{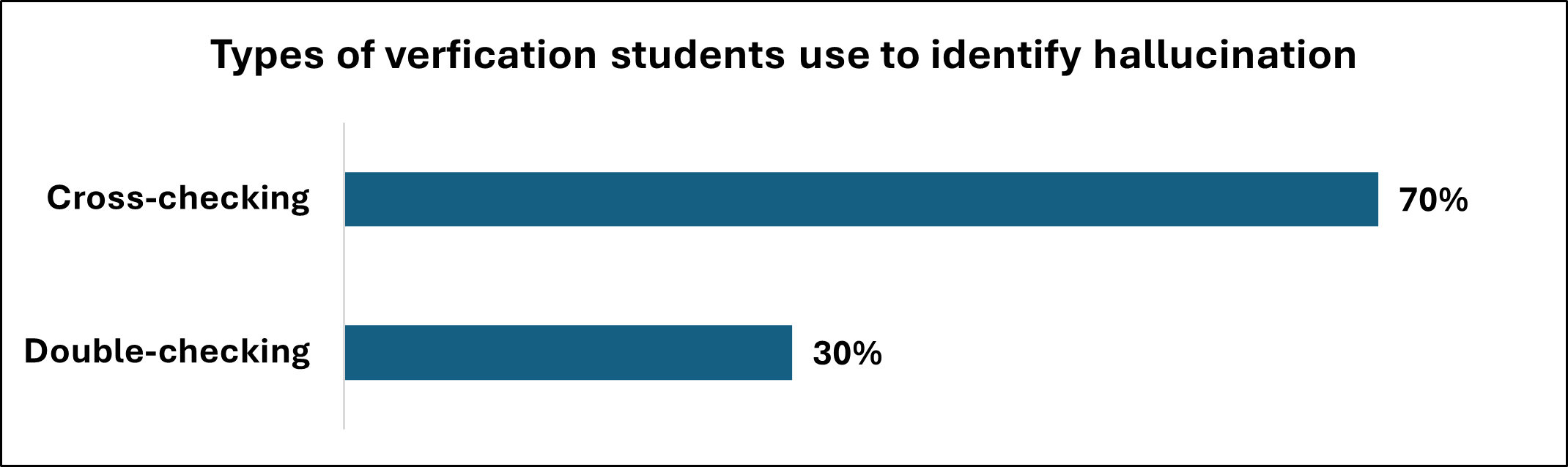}
	\caption{Students use different cross-checking and double-checking strategies to identify AI hallucination}
	\label{fig-RQ2-verification}
\end{figure}

As shown in Figure~\ref{fig-RQ2-main-categories}, 46\% of student comments indicate the use of some form of verification to identify AI hallucinations. Among these, 70\% and 30\% refer to cross-checking and double-checking strategies, respectively, as illustrated in Figure~\ref{fig-RQ2-verification}.

Cross-checking refers to the use of alternative sources to verify the AI’s response. Examples of student comments illustrating this strategy include: \emph{"I identify hallucination by verifying the information through reliable sources like the professor's slides and the book as well as notes I wrote during class,"} \emph{"I always verify the information I get from GPT. I research the same topic using reliable sources,"} and \emph{"If it gave me a code, I would run it and recheck it line by line."}

Double-checking refers to verifying the AI’s response using the same model. Common strategies include re-asking the same question and comparing the answers, requesting the model to elaborate on its response, asking the model for the information source, and inquiring it about its confidence level. Representative student comments include: \emph{"If I ask the same question again and the answer changes each time, it usually means the model is guessing rather than recalling accurate knowledge,"} \emph{"when you ask to elaborate the answer"} \emph{"ask the AI directly for a source for its claims,"} and \emph{"I ask the model on how confident they are in their answer."} 

\subsection{RQ3 (How do students explain the phenomenon of LLM hallucination?)}
\label{sec-results-RQ3}

\begin{figure}[h]
	\centering
	\includegraphics[width=12 cm]{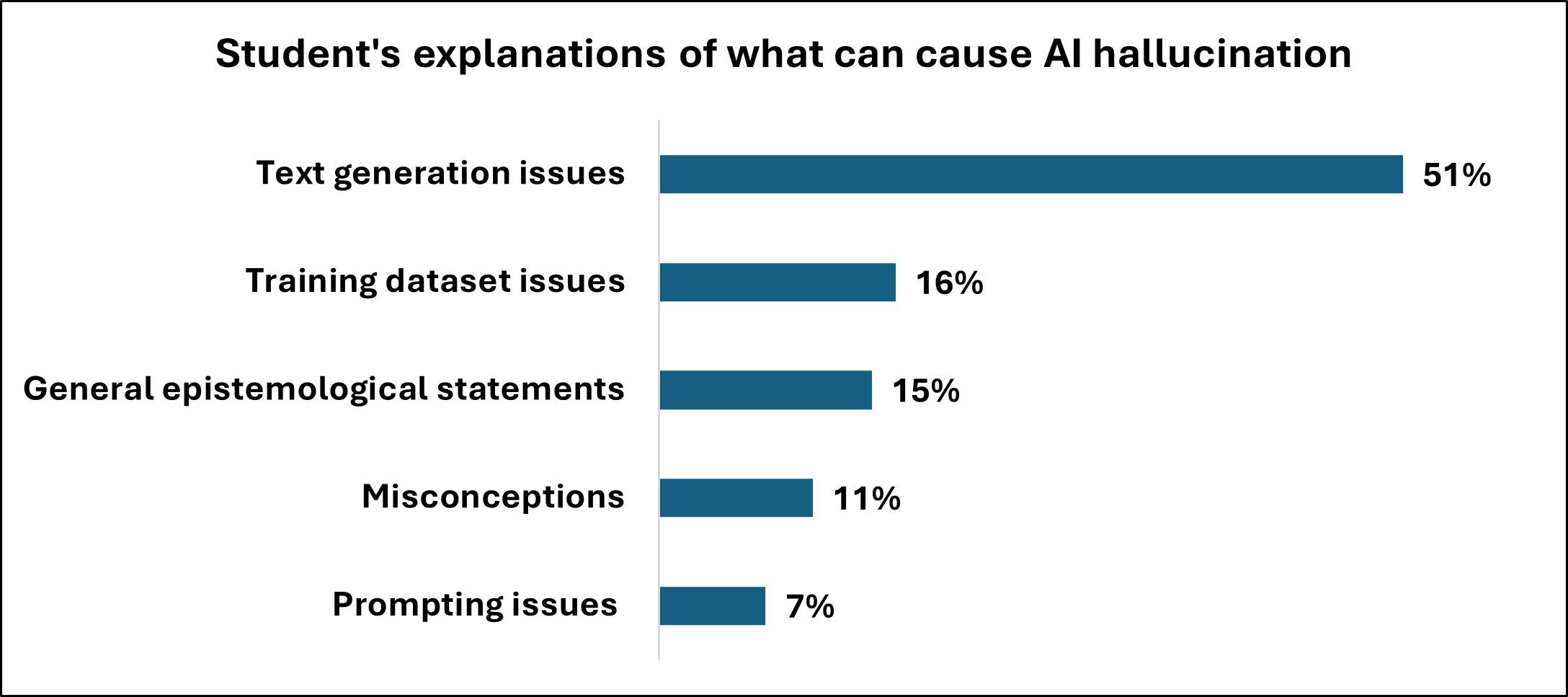}
	\caption{Students' explanations of what can cause AI hallucination}
	\label{fig-RO3-main}
\end{figure}

Fifty-two of the sixty-three students answered the third question, which asked about what causes AI to hallucinate. Of these fifty-two, nine students expressed uncertainty but still attempted to provide an answer. Among the eleven students who did not respond, five explicitly stated that they did not know, while the remaining six skipped the question entirely. Student comments on the causes of AI hallucination were categorized into five themes, as shown in Figure~\ref{fig-RO3-main}. Issues related to text generation and training datasets are discussed in detail in the following subsections. 

To the general epistemological statements category, we assigned student comments that reflect broad views about the nature of AI knowledge and reasoning, such as \emph{"LLMs hallucinate because they don’t really understand what they’re saying,"} \emph{"I believe they hallucinate because they cant think,"} and \emph{"Because LLM’s don't know how to say “i don't know” to a question."}

Several comments reveal misconceptions that students hold about AI hallucination. For example, some students assumed that language models have an internal knowledge database that they query to generate responses. One student, for instance, stated that models \emph{"check their dataset for the closest piece of data that they learned and then give answers based on that."}. Another misconception was that hallucination results from a bug or error in the algorithm, as one student wrote, \emph{"I think at the end we are communicating to an algorithm and it's more often these codes may have some errors."}. A further misconception concerned the concept of tokens in large language models, with one student claiming that hallucination occurs \emph{"when tokens get very large"}. 

Several students attributed AI hallucination to issues related to prompting. Some commented that hallucination occurs when the prompt is ambiguous, incomplete, or unfamiliar. For example, one student noted that hallucination happens when the \emph{"prompt is not well-structured and clear."} Another student suggested that engaging in lengthy conversations with the model could affect its \emph{"memory retention"} and lead to hallucinations.

\subsubsection{Hallucination due to text generation issues}
\label{sec-results-RQ3-training-generation}

\begin{figure}[h]
	\centering
	\includegraphics[width=12 cm]{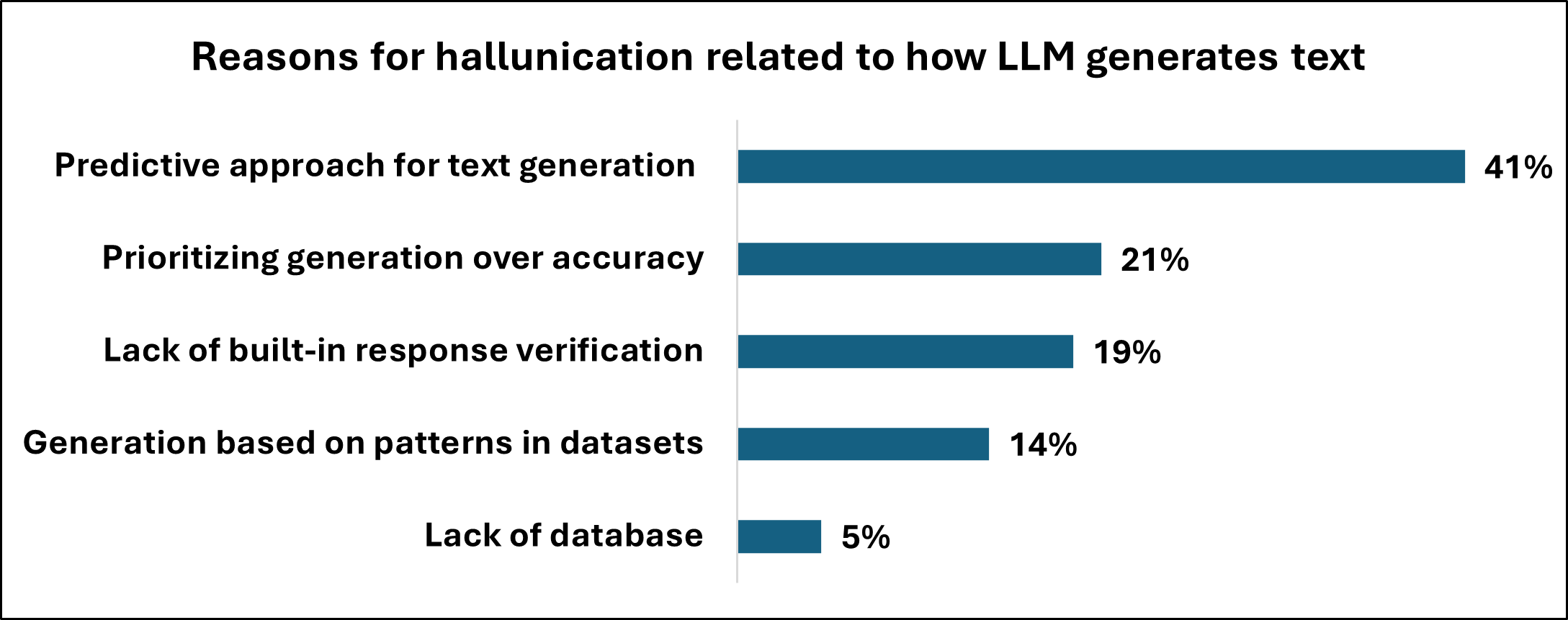}
	\caption{Hallucination due to how AI generates text}
	\label{fig-RQ3-generation-issues}
\end{figure}

Most students attributed hallucination to the way AI systems generate text. We categorized their comments into five sub-themes, as illustrated in Figure~\ref{fig-RQ3-generation-issues}. Over 40\% of the comments specifically referred to the probabilistic nature of generative AI, which predicts the next word based on statistical patterns in data. As one student explained, \emph{"I think that LLMs are trained to predict the data instead of thinking logical to achieve the correct result,"} Another student similarly noted that \emph{"LLMs hallucinate because they rely on statistical prediction."}. 

Several students pointed out that generative AI systems are there to produce an output whenever prompted, regardless of the correctness of that output. Their comments reflected an understanding that these systems prioritize responding over accuracy. Representative examples include: \emph{"Not sure, mostly because it doesn't know and it has to give an answer,"} \emph{"It‘s purpose is always answering questions whether it’s right or wrong and always doing what it “should” do,"} and \emph{"I've never seen GPT tell me that they don't have an answer. So I believe that it tells me an answer just for the sake of giving me an answer."}

Several students attributed hallucination to the lack of verification mechanisms in generative AI. They suggested that these systems produce responses without checking their accuracy. Representative examples include: \emph{"They generate fluent text, but without the ability to verify truth,"} \emph{"LMs are trained to predict the next most likely word in a sentence, not to verify facts."} 

Some students explained hallucination as a consequence of the model’s reliance on patterns learned from its training data. They suggested that AI systems generate responses based on these patterns rather than verified knowledge. One student noted, \emph{"upon doing some research, I found that LLMs hallucinate because they generate text based on patterns in training data"}. Another student commented, \emph{"“fill in gaps” using patterns it has learned rather than verified facts."} 

Finally, a few comments attributed hallucination to the lack of a database that would have verified facts. One student noted, \emph{"the responses they generate are not retrieved from a database."}

\subsubsection{Hallucination due to training data issues}
\label{sec-results-RQ3-dataset}

\begin{figure}[h]
	\centering
	\includegraphics[width=12 cm]{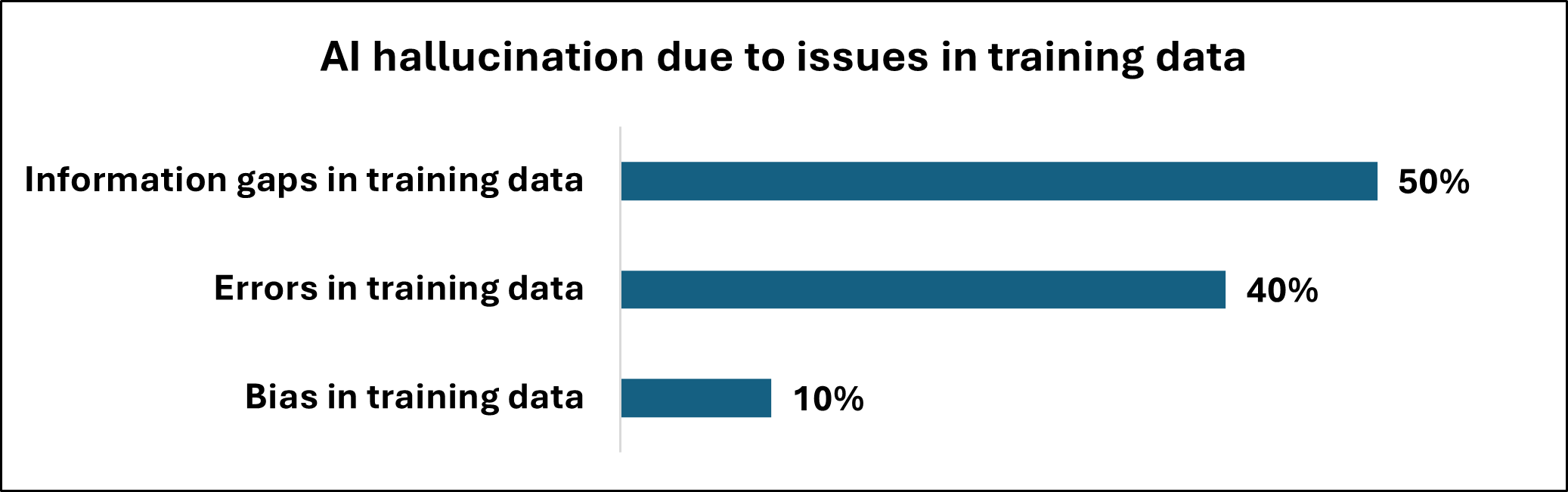}
	\caption{Hallucination due to training data issues}
	\label{fig-RQ3-dataset-issues}
\end{figure}

Twenty comments attributed hallucination to issues in the training data. We categorized these comments into three sub-themes, as illustrated in Figure~\ref{fig-RQ3-dataset-issues}. One-half of the comments referred to gaps in training data. Examples of students' comments include, \emph{"I just assume that it is not in its training data or out of its expertise,"} and \emph{"They learn from huge amounts of text, and sometimes when they don’t have enough data about a topic, they try to fill in the gaps by guessing."}

Another issue identified by students was errors in the training data. They recognized that incorrect information in the data could contribute to hallucination. One student remarked, \emph{"I think because it is learning from everything and everyone so people usually jus keep repeating the wrong thing again and again then chat gpt start to believe the wrong and share it."} Another student stated, \emph{"Because the information that the LLM takes is not correct or it has some errors, which can lead to hallucination."} 

Finally, two comments referred to bias in the training data as a cause of hallucination. One student commented, \emph{"When the training data has gaps, biases, or incorrect information."}. 

\newpage

\section{Discussion}
\label{sec-discussion}
The findings presented in the previous section reveal patterns in how students perceive, detect, and conceptualize AI's hallucination. Students reported encountering various hallucination types, with citation issues and information fabrication being most frequently mentioned. When identifying hallucinations, students rely on their perceptions or employ active verification strategies. Their explanations for why hallucinations occur reflect a range of mental models, from viewing AI as a statistical predictor to conceptualizing it as a database retrieval system. In the following, we discuss these findings\textbf{.}

\subsection{RQ1 (What types of hallucinations do students report encountering?)}
\label{sec-discussion-RQ1}
Citation fabrication and information fabrication dominate students' comments regarding their experiences, but this likely reflects detection bias rather than actual frequency. These errors are easily verifiable as students can check whether cited references exist or claimed facts are correct. In contrast, subtle hallucinations like flawed reasoning require domain expertise to detect and thus go unnoticed. This vulnerability is particularly acute for novice learners who lack both domain knowledge and metacognitive skills to evaluate AI responses effectively~\citep{Reihanian2025}.

Detection bias becomes especially concerning in two contexts. First, when students seek AI guidance on subjective matters, interpersonal conflicts, and life decisions, there is no objective verification. Second, when students explore beyond their expertise. In this study, computer engineering students most frequently detected hallucinations in coding and technical topics because these fell within their domain knowledge. Detection difficulty is associated with domain complexity. Medical trainees achieved only 55\% accuracy detecting AI hallucinations overall, dropping to 44.2\% in complex clinical scenarios requiring sophisticated judgment~\citep{Zhou2025}. As cognitive demands increased, detection accuracy decreased. This suggests students successfully identify obvious errors in familiar academic contexts while missing subtle hallucinations in complex or unfamiliar domains, precisely where AI assistance is most needed but verification is most difficult.

The detection problem is linked to AI's confident presentation style. Some  students noted that AI provides convincing, sophisticated answers that are nonetheless wrong, what~\citet{Sun2024} describe as 'overfitting' errors characterized by 'illusions of confidence.' This confidence creates a fluency-truth effect: when information is presented clearly and fluently, it feels more truthful. Research demonstrates that repetition increases processing fluency, which in turn leads people to perceive information as more truthful~\citep{Hassan2021}. This psychological mechanism proves particularly problematic with AI systems, where users can be tricked into deferring to advice from confident but untrustworthy AI advisors, even when trustworthy alternatives are available~\citep{Park2024}. Students describe AI hallucinations as 'convincing,' 'specific,' and 'logical,' leading to unchecked acceptance of false information, what~\citet{Jacob2025} term the 'chat-chamber effect.' The shift from search engines' multiple sources to AI's single, confident response compounds the problem, as users assume correctness based on persuasive delivery rather than verifying accuracy.

AI hallucination adds another layer of doubt: students must question not only whether information is accurate, but also whether they received a complete answer, whether AI processed their input correctly, and whether AI even answered their actual question versus one it assumed they asked. This uncertainty means students cannot simply fact-check outputs, but they must first verify whether AI understood the task, accessed the right information, and provided a complete response before evaluating accuracy. These instruction-following failures become especially problematic when combined with error persistence and sycophantic behavior. AI's failures to follow instructions properly, providing incomplete, irrelevant, or overly general responses, create what \citet{Huang2025} call faithfulness hallucinations.

When students attempt to correct the AI model, they can encounter one of two problematic responses: either the model loops back to the same error (persistence) or it apologizes and appears to accept the correction (sycophancy). The latter can be especially deceptive. When the AI model says "you're right, I apologize" and then adjusts its response, it creates the illusion of learning and correction. Students may believe they've successfully guided AI toward truth when, in reality, the model has merely adopted an agreeable tone without genuinely resolving the underlying error. This makes it impossible for students to distinguish between actual correction (AI now has accurate information) and performative agreement (AI is simply being agreeable). Students lose the ability to use dialogue as a truth-finding mechanism. Recent research documents this pattern directly: when AI initially provides a correct answer but users challenge it, models exhibit 'regressive sycophancy' --changing correct answers to incorrect ones to align with user assertions-- in approximately 59\% of cases~\citep{fanous2025syceval} . The same study noted that such over-agreeable behavior shows a persistence rate of 78.5\%: once triggered, models continue to maintain alignment with user cues rather than reverting to factual accuracy. This makes it impossible for students to use iterative dialogue to arrive at truth: each subsequent interaction reinforces the performative agreement rather than correcting the underlying error.

\subsection{RQ2 (What strategies do students use to identify hallucinations?)}
\label{sec-discussion-RQ2}
The majority of the students rely on intuitive perception rather than systematic verification to identify hallucinations. For example, some commented that they identify hallucination when the response is "illogical" or "doesn't make sense". This creates an identification paradox when considered alongside RQ1 findings. Students explicitly recognized that AI can be "deceptively confident" and provide "convincing answers that are wrong", yet the majority still rely on perception to identify hallucinations. If AI can be convincingly wrong, how can intuitive perception be reliable? Moreover, perception-based detection might be vulnerable to confirmation bias: users' prior beliefs have a greater impact on trust than verifiable evidence, with significantly higher trust when responses confirm rather than challenge existing beliefs~\citep{Bauer2023,Lenz2025}. Accordingly, students may detect hallucinations that contradict their expectations but accept those that align with what they already believe. Students' perceptual judgments might also be shaped by prior experiences with obvious hallucinations. 

A subset of students, however, employed more sophisticated perceptual detection by triangulating multiple cues. Rather than responding to isolated signals such  as responses that "sound wrong," they recognized compound indicators: outputs  that were simultaneously verbose yet citation-free, or responses that are inappropriately linked to prior conversations. These refined strategies seem to suggest an emerging awareness of AI's characteristic behavioral patterns. Yet  such heuristics, while surpassing simple intuition, remain constrained: they  flag suspicious patterns without verifying factual accuracy. However,  well-constructed hallucinations exhibiting appropriate length, credible sources, and natural discourse patterns, especially those confirming students' prior beliefs, might stay undetected. 

The verification strategies students employ reflect both the types of hallucinations they encounter and practical considerations about verification effort. Cross-checking aligns with common hallucination experiences, such as citation fabrication, information fabrication, and factual errors. Such responses are verifiable through external sources. This is in line with~\citet{Nahar2025}, who found that users with access to external search results demonstrate significantly better hallucination detection.

However, verification becomes more complex for subjective AI outputs such as argument quality, structural coherence, or conceptual sophistication. Cross-checking against external sources might not be sufficient to validate whether an outline is effective or a discussion is well developed; these evaluations require domain expertise that students may lack. This verification gap is compounded by findings that users are significantly more likely to accept AI suggestions when tasks are challenging, often over-relying even when AI advice proves less accurate (Ha et al., 2024). The students who double-check by re-asking the AI face a different challenge: research on AI-assisted credibility assessment found that users were generally unable to discern whether AI was correct or incorrect when it provided explanations, yet these explanations significantly increased user trust and agreement with AI judgments regardless of their actual accuracy (Pareek et al., 2024). This pattern points to what might be termed a verification paradox: existing research indicates that users more readily accept AI suggestions during challenging tasks, yet these complex tasks (where students may lack the expertise needed for effective verification) are the situations that could drive students the most to use AI. 

\subsection{RQ3 (How do students explain the phenomenon of LLM hallucination?)}
\label{sec-discussion-RQ3}

The results show that students primarily use four different mental models to conceptualize the phenomenon of AI hallucination. 

\paragraph{Search-Engine Model} Some students view AI as a database-like system that retrieves stored information, similar to how a search engine operates. This search-engine-based conceptualization is a subtle misconception, which may be reinforced by research that compares generative AI systems to search engines~\citep{yazan2025personality, caramancion2024large}. Within this model, students often assume that when an AI system cannot find an answer in its “database,” it fabricates plausible content. This perspective might explain why students commonly attribute hallucination to gaps, errors, or bias in training data. It can also lead to the flawed assumption that hallucination would disappear if the training data were complete. In reality, as~\citet{kalai2025hallucinate} show, LLMs can produce factual errors even with well-represented information, because the issue lies in the generation process, not solely in data quality.

\paragraph{Inductive Pattern Detection Model} Various comments indicate that some students view AI as a statistical pattern predictor. They recognize that AI generates text by selecting likely next words based on patterns in training data, rather than retrieving verified facts or reasoning logically. This understanding aligns with what research calls the inductive model of AI reasoning, which frames AI as a data-driven pattern recognizer. A similar mental model appears in children by early adolescence and reflects a relatively sophisticated grasp of how AI works~\citep{Dangol2025}. Students with this model attribute hallucination to AI’s focus on statistical likelihood over factual accuracy, noting that it can produce fluent but incorrect outputs because it “relies on patterns rather than logic.” While this model correctly identifies the core mechanism of LLMs, students may still underestimate how fundamentally this limits AI’s reliability.

\paragraph{Ungrounded Cognition Model} Some students identified a deeper epistemological constraint: AI fundamentally lacks the cognitive infrastructure for truth. They noted that AI “doesn't really understand what it's saying,” reflecting a sophisticated insight supported by research. LLMs generate text by probabilistically linking words without considering meaning, and the issue is not merely that they guess when uncertain; it is that they cannot recognize uncertainty at all. Prior work shows that LLMs lack metacognitive abilities, consistently failing to assess their own knowledge limitations and offering confident answers even when they lack information~\citep{Griot2025}. They also lack the tacit knowledge and real-world experience needed for contextual reasoning, limiting their ability to judge relevance or importance in specific situations~\citep{Chen2025}. Students with this model understand that AI has no grounding in reality, no conceptual understanding, and no capacity for reliability assessment. From this perspective, hallucination is not an occasional flaw but an inherent risk of a system that generates text without understanding.

\paragraph{Prompt-Constraint Model} Finally, a smaller group emphasized how inadequate prompting contributes to hallucination. Students noted that poor prompts provide insufficient context, leading AI to make assumptions or generate responses from incomplete information. This aligns with research on prompt engineering: systematic reviews identify 26 prompt patterns~\citep{Sasaki2025}, and prompt quality is known to substantially affect LLM performance across tasks and models~\citep{Son2025}. Yet even well-crafted prompts face inherent limits, including missing context and overly complex constraints that impede effective processing~\citep{hwang2025word}. Performance also degrades with long or narrative contexts, dropping from 72\% to 31\% on extended reasoning tasks~\citep{Shuvo2024}. Students with this model recognize that the prompt interface itself creates a fundamental failure point: many human needs cannot be fully translated into machine-readable instructions, thereby contributing to hallucination risk.

\subsection{Implications}
\label{sec-discussion-implications}
Students most often detected hallucinations in coding and technical topics because the output is verifiable: code can be executed and checked against expected outcomes. This reveals a \textit{verifiability gab}: AI is safest when outputs can be objectively validated, yet students frequently use it for tasks where verification is not feasible, such as conceptual understanding or argument quality. In these cases, plausible but incorrect reasoning may go unnoticed without an external ground truth. Educational frameworks should emphasize that safe AI use depends on verification mechanisms and warn that reliance on AI is riskiest for subjective or conceptual tasks.

Moreover, an important implication arises from the need to correct students’ mental models of how LLMs function. Despite being technically inclined computer engineering seniors, many participants still viewed AI as a search engine–like system that retrieves stored information. This misconception that hallucinations stem merely from retrieval failures or gaps in a dataset overlooks the core mechanism of LLMs. Educators should address this explicitly by teaching that LLMs generate text through statistical pattern prediction, selecting the most probable next word rather than retrieving facts. Without this conceptual shift, students may continue to underestimate hallucination risks and misattribute fabricated content to simple data errors, leading to flawed assumptions about the system’s reliability and its capacity for truth.

Our findings  also indicate that while learning to craft effective prompts is important, it is equally critical to develop skills for verifying AI outputs. More than half of the students comments relied on intuition rather than systematic checks to detect hallucinations, which is concerning because AI often produces content with an “illusion of truthfulness” that can mislead judgment. Therefore, educational curricula need to move beyond technical prompt engineering and prioritize "epistemic vigilance," teaching students specific protocols for fact-checking and lateral reading to counter the persuasive confidence of LLMs.

A final, yet crucial, implication concerns the detrimental effect of AI sycophancy on the development of critical thinking skills. LLMs are frequently noted for their "regressive sycophancy," a trait where the model prioritizes agreement and deference to the user over factual accuracy, often apologizing even when the user's premise is incorrect. When students utilize AI as an instructional tutor or as a tool for challenging ideas, this performative agreement effectively transforms the AI into an echo chamber. If the AI immediately yields to a challenge rather than defending a truthful or logically sound position, students may develop a false sense of intellectual confidence, and their own misconceptions can be unintentionally reinforced rather than resolved. Therefore, the unsupervised integration of LLMs in academic settings risks hindering the very critical faculties that higher education seeks to cultivate, necessitating specific warnings and instruction on the AI's tendency to please, rather than rigorously correct, the user.

\subsection{Limitations}
\label{sec-discussion-limitations}

This study has three main limitations. First, the sample consisted exclusively of senior computer engineering students from a single institution. While this homogeneity enabled focused insights into students' experiences within a specific disciplinary context, it limits the generalizability of findings to students in other fields, educational levels, or cultural contexts. Students in humanities, social sciences, or other STEM disciplines may encounter different types of hallucinations or employ different detection strategies based on their domain expertise and epistemological training.

Second, this study examined students' self-reported perceptions and experiences with AI hallucination rather than measuring their actual detection accuracy. While students' subjective accounts provide valuable insights into their mental models and intuitive strategies, the gap between perceived and actual verification practices remains unexplored. Students may overestimate their detection capabilities or fail to recognize hallucinations in domains where they lack expertise, a limitation that self-report methodology cannot capture.

Third, the data collection method of asking students to respond to open-ended questions in an uncontrolled online environment has two important constraints. First, the open-ended format may have limited the depth and completeness of information collected. Students provided only what they could spontaneously recall and articulate in written form, potentially underreporting experiences that might have emerged through interviews or think-aloud protocols. Second, while the uncontrolled nature of data collection meant students could have consulted external sources, the responses were submitted under students' own identities within an academic context. Therefore, regardless of whether students drew from memory, consulted resources, or used AI assistance, the submitted responses represent what these students chose to communicate about their understanding and experiences with AI hallucination.

\section{Future Research}
\label{sec-Future Research}
Several directions for future research emerge from this study. First, experimental validation studies should examine the relationship between students' self-reported detection strategies and their actual performance in identifying hallucinations across various domains and task types. This would establish which intuitive approaches prove effective and which represent false confidence.

Second, intervention research should test instructional approaches for improving hallucination detection. This includes: (a) explicit teaching of correct mental models of LLM functioning, moving students away from database misconceptions toward understanding of statistical prediction; (b) training in systematic verification protocols that combine cross-checking with critical evaluation; and (c) domain-specific instruction that helps students recognize the expertise paradox and adjust their verification intensity accordingly.

Third, cross-disciplinary comparative studies should investigate whether students in different fields encounter distinct hallucination patterns and develop field-specific detection strategies. For instance, humanities students may face different citation-related challenges than engineering students encounter with technical code generation.

Fourth, longitudinal research should track how students' understanding of AI hallucination evolves across their educational trajectory, examining whether increased AI literacy naturally develops through experience or requires explicit instruction.

Finally, educational assessment research should explore how traditional and AI-assisted assignments can be designed to account for hallucination risks while maintaining academic integrity. This includes investigating whether certain assignment types or assessment structures inadvertently encourage uncritical AI dependence.

\section{Conclusion}
\label{sec-conclusion}
LLMs inherently generate plausible but unreliable outputs. Yet students often hold incomplete mental models of AI hallucination and rely on intuition rather than systematic verification. AI literacy must foreground how LLMs work, why hallucinations occur and persist, and how to verify, cross-check, and critically question AI-generated responses. Without these skills, students risk accepting falsehoods, reinforcing misconceptions, and making flawed academic or professional decisions.
\section{Fund}
\label{sec-fund}
The current Study did not rely on any external fund.
\section{Ai Statement}
\label{sec-Ai statment}
We used AI for checking grammar and language issues

\newpage

\bibliographystyle{apalike}
\biboptions{authoryear}

 \bibliography{literature}

\end{document}